\documentclass[12pt,english]{article}
\usepackage{graphicx}
\begin{document}
\title{A note on the Casimir energy of a massive scalar field in  positive
curvature space}
\author{E. Elizalde\footnote{e-mail: elizalde@ieec.fcr.es}\\
 Consejo Superior de Investigaciones Cient\'{\i}ficas \\
 Institut d'Estudis Espacials de Catalunya (IEEC/CSIC) \\
 Edifici Nexus, Gran Capit\`{a} 2-4, 08034 Barcelona, Spain; \& \\
 Departament d'Estructura i Constituents de la Mat\`{e}ria \\
 Facultat de F\'{\i}sica, Universitat de Barcelona \\
  Diagonal 647, 08028 Barcelona, Spain \\
A. C. Tort%
\footnote{e-mail: tort@if.ufrj.br. }\\
 Departamento de F\'{\i}sica Te\'{o}rica - Instituto de F\'{\i}sica
\\
Universidade Federal do Rio de Janeiro\\
Caixa Postal 68.528; CEP 21941-972 Rio de Janeiro, Brazil}
\maketitle
\begin{abstract}
\noindent We re-evaluate the zero point Casimir energy for the
case of a massive scalar field in $\mathbf{R}^{1}\times
\mathbf{S}^{3}$ space, allowing also for deviations from the
standard conformal value $\xi =1/6$, by means of zero temperature
zeta function techniques. We show that for the problem at hand
this approach is equivalent to the high temperature regularization
of the vacuum energy, as conjectured in a previous publication.
Two different, albeit equally valid, ways of doing the analytic
continuation are described.
\bigskip \bigskip

\vfill
\end{abstract}
%
%
In recent papers, the vacuum and finite temperature energies for
massless and  massive scalar fields in positive curvature spaces,
as $\mathbf{S}^3$, were considered, with special emphasis being
put on the analysis of  entropy bounds corresponding to those
cases [see Refs. \cite{Breviketal2002} and
\cite{Elizalde&Tort2003}, respectively]. Specifically, in Ref.
\cite{Elizalde&Tort2003}, dealing with the massive case and
arbitrary coupling, the vacuum (Casimir) energy and the finite
temperature energy were obtained through the application of a
generalized zeta function technique \cite{Elizalde94}, which
provides a neat formal separation of the logarithm of the
partition function into the non-thermal and thermal sectors. This
approach holds generally and, in particular, it does for
$\textbf{S}^1\times \textbf{S}^d$ spaces. For the special case
$d=3$, the result for the logarithm of the partition function can
be combined with the Abel-Plana rescaled sum formula
\cite{Erdelyietal1953}, leading to \cite{Elizalde&Tort2003}
\begin{eqnarray}\label{logZ}
\log Z\left(\beta \right)&=&-\frac{\beta }{2r}\sum _{n=1}^{\infty }n^{2}
\left(n^{2}+\mu _{eff}^{2}\right)^{1/2}+\frac{r\mu _{eff}^{2}}{\beta }
\sum _{n=1}^{\infty }\frac{1}{n^{2}}K_{2}\left(m_{eff}\beta n\right)
 \nonumber \\ &+&\frac{\mu _{eff}^{2}\beta }{\left(2\pi \right)^{2}r}\,
\sum _{n=1}^{\infty }\frac{1}{n^{2}}K_{2}\left(2\pi n\mu
_{eff}\right) -\frac{\mu _{eff}^{3}\beta }{2\pi r}\sum
_{n=1}^{\infty }\frac{1}{n}K_{3} \left(2\pi n\mu _{eff}\right).
\end{eqnarray}
From this result the present authors conjectured that the
\emph{renormalized} vacuum energy could be inferred as
\begin{equation}\label{Representation1}
E_0=-\frac{\mu^2_{eff}}{\left(2\pi\right)^2r}\sum_{n=1}^\infty
\frac{1}{n^2}K_2\left(2\pi
n\mu_{eff}\right)+\frac{\mu^3_{eff}}{2\pi r}
\sum_{n=1}^\infty\frac{1}{n}K_3\left(2\pi n\mu_{eff}\right),
\end{equation}
where $r$ is the radius of $\mathbf{S}^3$, $\beta$ is the
reciprocal of the temperature, and the parameter $\mu_{eff}$, to
be defined below, plays the role of an `effective mass'. The
conjecture stems from the fact that in Eq.~(\ref{logZ}), the terms
linear in $\beta$ give rise to temperature independent terms when
we calculate basic thermodynamics quantities such as the free
energy and the energy. Also, in the very-high temperature limit,
it is physically plausible to expect  the Stefan-Boltzmann term
---which is contained in the second term in Eq.~(\ref{logZ})--- to be
the only surviving one. It must be stressed that the renormalized
vacuum energy was calculated with a representation derived from
the finite temperature sector of the problem at hand, an approach
advocated in \cite{KE1996} and made easier, in the present case,
by the absence of the zero mode.

In this note, we wish to show that the above-mentioned conjecture
holds indeed true and that the standard zero-temperature zeta
function regularization procedure (as prescribed in
\cite{Elizalde94}, for instance)  works very nicely, yielding the
correct vacuum energy for the non-conformal case. In what follows
we will also show that the zeta function technique is free from
the disturbing difficulties met by Green function techniques
\cite{Breviketal2002} ---and also by other techniques--- when
applied to the same problem. In all evaluations natural units will
be employed. 

The vacuum energy for a massive scalar field in $\mathbf{S}^3$ and
arbitrary conformal parameter $\xi$, which can also be read out
from Eq.~(\ref{logZ}), is given by
\begin{equation}
E_0=\frac{1}{2r}\sum_{\ell=0}^\infty D_{\ell}M_{\ell},
\end{equation}
where
\begin{equation}
M_{\ell }^{2}:=\left(\ell +1\right)^{2}+\mu _{eff}^{2}
\end{equation}
and the dimensionless parameter $\mu _{eff}^{2}$ is defined as
\begin{equation}\label{effective mass}
\mu _{eff}^{2}=\mu ^{2}+\chi -1,
\end{equation}
where $\mu:=mr$, $m$ being the mass of an elementary excitation of
the scalar field, $r$ the radius of $\mathbf{S}^3$, and
$\chi:=\xi{\cal R}r^2$; ${\cal R}$ is the Ricci curvature scalar.
It is easy to see that the parameter $\mu_{eff}$ plays a role
similar to that of an effective mass. Note, moreover, that $\mu
_{eff}^{2}$ and $\chi$ are real numbers and that $\mu ^{2}\geq 0$.
The degeneracy factor is $D_{\ell }=\left(\ell +1\right)^{2}$.
Hence, for a massive scalar field the Casimir energy is formally
given by
\begin{equation}\label{zero-point energy}
E_{0}\left(\mu _{eff}^{2}\right)=\frac{1}{2r}\sum _{\ell
=0}^{\infty } \left(\ell +1\right)^{2}\sqrt{\left(\ell
+1\right)^{2}+\mu _{eff}^{2}},
\end{equation}
an expression which needs to be conveniently regularized. This
will be always done in what follows by using zeta function
techniques. Thus, the standard conformal case corresponds to the
values $\mu ^{2}=0$ and $\chi =1$ ($\xi =1/6$, ${\cal R}=6/r^2$)
and, using the zeta function of the Hamiltonian operator, we
obtain in this simple case
\begin{equation}
E_{0}\left(\mu ^{2}=0,\chi =1\right)=\frac{1}{2r}\zeta \left(-3\right)=
\frac{1}{240r}.
\end{equation}
Our problem now is, however, to find an analytical continuation in
terms of the zeta function for the more difficult series in
Eq.~(\ref{zero-point energy}) and give thereby both a mathematical
and a physical meaning to it.

Starting from Eq.~(\ref{zero-point energy}) we can
straightforwardly write:
\begin{equation}
\left. \sum _{\ell=0}^{\infty
}\left(\ell+1\right)^{2}\left[\left(\ell+1\right)^{2}+\mu_{eff}^2
\right]^{-s} \right|_{s=-\frac{1}{2}}=F(-3/2,\mu_{eff}^2) -
\mu_{eff}^2F(-1/2,\mu_{eff}^2), \label{soli1}
\end{equation}
being, by definition,
\begin{equation}
F(s,\mu_{eff}^2) :=\sum _{\ell=0}^{\infty }
\left[\left(\ell+1\right)^{2}+\mu_{eff}^2 \right]^{-s}
\end{equation}
 or alternatively, after simple manipulations,
\begin{equation}
\left.\sum _{\ell=0}^{\infty
}\left(\ell+1\right)^{2}\left[\left(\ell+1\right)^{2}+\mu_{eff}^2
\right]^{-s} \right|_{s= -\frac{1}{2}}=\left.\frac{1}{1-s}
\frac{\partial }{\partial t}\,  G(s-1,t,\mu_{eff}^2)
\right|_{t=1;s=-\frac{1}{2}}, \label{soli2}
\end{equation}
with
\begin{equation}
G(s,t,\mu_{eff}^2) :=\sum _{\ell=0}^{\infty }
\left[\left(\ell+1\right)^{2}t+\mu_{eff}^2 \right]^{-s}
\end{equation}
 It is not difficult to prove that both expressions,
(\ref{soli1}) and (\ref{soli2}), lead to the same result, as
described below.

 The sum over $\ell$ can be
identified as an Epstein series \cite{Elizalde94,ER89a}
\begin{equation}
E_{N}^{c}\left(z;\vec{a};\vec{c}\right):=\sum _{n_{1}
\ldots n_{N}=0}^{\infty }\left[a_{1}\left(n_{1}+c_{1}\right)^{2}
+\cdots +a_{N}\left(n_{N}+c_{N}\right)^{2}+c\right]^{-z}
\end{equation}
where $a_{i}>0,c_{i}>0$ and $c>0$. The analytical continuation in
the special case $N=1$ is given by
\begin{eqnarray}
E_{N}^{c}\left(z;\vec{a};\vec{c}\right): & = & \frac{c^{-z}}{\Gamma
\left(z\right)}\sum _{p=0}^{\infty }\frac{\left(-1\right)^{p}\Gamma
\left(z+p\right)}{p!}\left(\frac{a_{1}}{c}\right)^{p}\zeta \left(-2p,c_{1}
\right)+\frac{c^{\frac{1}{2}-z}}{2}\sqrt{\frac{\pi }{c_{1}}}\,
\frac{\Gamma \left(z-\frac{1}{2}\right)}{\Gamma \left(z\right)}\nonumber \\
 & + & \frac{2\pi ^{z}}{\Gamma \left(z\right)}\cos \left(2\pi c_{1}\right)
 a_{1}^{-\frac{z}{2}-\frac{1}{4}}c^{-\frac{z}{2}+\frac{1}{4}}
 \sum _{n_{1}=0}^{\infty }n_{1}^{z-\frac{1}{2}}\, K_{z-\frac{1}{2}}\left(2
 \pi n_{1}\sqrt{\frac{c}{a_{1}}}\right).
\end{eqnarray}
In our case, after the identifications:
$a_{1}=1,c_{1}=1,z=s-1$, and $c\equiv \mu_{eff}^2t^{-1}$,
we have
\begin{eqnarray}
E_{1}^{\mu_{eff}^2t^{-1}}\left(s-1;1;1\right) & = & \sum
_{p=0}^{\infty }\left[\left(p+1\right)^{2}+\mu_{eff}^2t^{-1}
\right]^{1-s} \nonumber \\&& \hspace*{-36mm} =
\frac{\left(\frac{\mu_{eff}^2}{t}\right)^{1-s}}{\Gamma
\left(s-1\right)}\sum _{p=0}^{\infty
}\frac{\left(-1\right)^{p}\Gamma
\left(s-1+p\right)}{p!}\left(\frac{t}{\mu_{eff}^2}\right)^{p}\zeta
\left(-2p,1\right) +\frac{\sqrt{\pi
}\left(\frac{\mu_{eff}^2}{t}\right)^{\frac{3}{2}-s}}{2} \nonumber
\\ \times\frac{\Gamma \left(s-\frac{3}{2}\right)}{\Gamma
\left(s-1\right)} & + &\frac{2\pi^{s-1}}{\Gamma
\left(s-1\right)}\left(\frac{\mu_{eff}^2}{t}
\right)^{-\frac{s}{2}+\frac{3}{4}} \sum _{n_{1}=0}^{\infty
}n_{1}^{s-\frac{3}{2}}\,K_{s-\frac{3}{2}}\left(2\pi
n_{1}\frac{\mu_{eff}}{t^{1/2}}\right).
\end{eqnarray}
Using now the first of the two alternatives, Eq.~(\ref{soli1}),
doing some calculations we obtain for the vacuum energy:
\begin{equation}
E_0=-\frac{\mu^4_{eff}}{16r}\Gamma\left(-2\right)+\frac{3\mu^2_{eff}}{4\pi^2
r} \sum_{n=1}^\infty\frac{1}{n^2}K_2\left(2\pi n\mu_{eff}\right)
+\frac{\mu^3_{eff}}{2\pi
r}\sum_{n=1}^\infty\frac{1}{n}K_1\left(2\pi
n\mu_{eff}\right).\label{Sumpri}
\end{equation}
Alternatively, by using Eq.~(\ref{soli2}), that is, taking the
derivative with respect to $t$ and setting $t=1$ and $s=-1/2$ we
arrive, after some manipulations, at the following (differently
looking) expression for the vacuum energy
\begin{equation}\label{SumOne}
E_0=-\frac{\mu^4_{eff}}{16r}\Gamma\left(-2\right)+\frac{\mu^2_{eff}}{4\pi^2
r} \sum_{n=1}^\infty\frac{1}{n^2}K_2\left(2\pi
n\mu_{eff}\right) -\frac{\mu^3_{eff}}{2\pi
r}\sum_{n=1}^\infty\frac{1}{n}\frac{\partial} {\partial
u}K_2\left(u=2\pi n\mu_{eff}\right).
\end{equation}
The divergent term in Eqs.~(\ref{Sumpri}) and (\ref{SumOne}) may
be taken care of with the minimal subtraction scheme
\cite{Blauetal1988}, what means to take into account the principal
part of $\Gamma\left(-2\right)=\psi\left(3\right)/2$. However,
this procedure would imply, in our case, to keep a quartic term in
$\mu_{eff}$ that would spoil the behavior of the vacuum energy in
the classical limit, where the vacuum oscillations are known to
vanish. As a consequence, old-style physical renormalization
immediately prescribes that this term must be simply discarded,
what we will do from now on, without too much ado.

We can now further simplify Eq.~(\ref{SumOne}), by making use of
the recursion relation \cite{Grad94}
\[
-2\frac{dK_\nu\left(z\right)}{dz}=K_{\nu-1}\left(z\right)+K_{\nu+1}
\left(z\right),
\]
which allows us to write
\begin{eqnarray}
E_0&=&\frac{\mu^2_{eff}}{4\pi^2 r}\sum_{n=1}^\infty
\frac{1}{n^2}K_2\left(2\pi n\mu_{eff}\right) \nonumber \\
&+&\frac{\mu^3_{eff}}{4\pi
r}\sum_{n=1}^\infty\frac{1}{n}K_1\left(2\pi n
\mu_{eff}\right)+\frac{\mu^3_{eff}}{4\pi
r}\sum_{n=1}^\infty\frac{1}{n} K_3\left(2\pi
n\mu_{eff}\right).\label{Representation2}
\end{eqnarray}
If we now make use of this other recursion relation:
\[
K_{\nu-1}\left(z\right)-K_{\nu+1}\left(z\right)=-\frac{2\nu}{z}K_\nu
\left(z\right),
\]
we easily obtain {\it both} Eq.~(\ref{Representation1}),
conjectured in a former paper, and also Eq.~(\ref{Sumpri}), which
are completely equivalent, by the same recursive formulas. We have
thus shown that all these approaches are perfectly equivalent.

When we take the limit $\mu_{eff}^2\to 0$,
Eqs.~(\ref{Representation1}) and (\ref{Representation2}) yield the
well known result $E_0\approx 1/240$. Moreover, if we make use of
the appropriate small argument expansion of the Bessel functions
of the second kind, we obtain
\begin{equation}
E_0\approx \frac{1}{240 r}-\frac{\mu^2_{eff}}{48
r}-\frac{1}{2}
\left[\frac{1}{8r}+\frac{1}{16r}\left(-\frac{3}{2}+2
\gamma_E\right)\right]\mu^4_{eff}.
\end{equation}
On the other hand, if we take the opposite limit, $\mu_{eff}^2\to
\infty$, the vacuum energy given by Eq.~(\ref{Representation1})
---or (\ref{Sumpri}) or (\ref{Representation2})--- behaves in the
way that one would normally expect of constrained zero-point
oscillations of a massive quantum field: the vacuum energy goes to
zero in an exponential way.

 In Ref. \cite{Breviketal2002} the case of a massless
scalar was considered. In order to obtain the vacuum energy, the
formalism due to Kantowski and Milton \cite{Kantowski&Milton87},
which relies on Green function techniques, was successfully
employed. However, when the same techniques were applied to the
conformal symmetry breaking case, \emph{ab initio} unsurmountable
difficulties were met. The finite temperature energy evaluation is
also plagued by difficulties. The good news are that none of these
problems are present when we have recourse to the generalized zeta
function procedure, as we have here shown.
\section*{Acknowledgments}
A.C.T. wishes to acknowledge the kind hospitality of the Institut
d'Estudis Espacials de Catalunya (IEEC/CSIC) and of the
Universitat de Barcelona, Departament d'Estructura i
Cons\-tituents de la Mat\`{e}ria, where the present work was
began. The investigation of E.E. has been supported by DGI/SGPI
(Spain), project BFM2000-0810, and by CIRIT (Generalitat de
Catalunya), contract 1999SGR-00257.
\bigskip \bigskip
%

%
\end{document}